\documentclass[pre,showpacs,twocolumn]{revtex4}
\input{psfig.sty}
\begin{document}

\title{
  Long range frustration in finite connectivity spin glasses:
  A mean field theory and its 
  application to the random $K$-satisfiability problem
}

\author{
  Haijun Zhou
}

\affiliation{
  Max-Planck-Institute of Colloids and Interfaces, 14424 Potsdam, Germany
}

\affiliation{
  Institute of Theoretical Physics, the Chinese Academy of Sciences,
  Beijing 100080, China
}


\begin{abstract}
  A mean field theory of long range frustration is constructed for
  spin glass systems with quenched randomness of vertex--vertex
  connections and of spin--spin
  coupling strengths. This theory is applied to a spin glass
  model of the random $K$-satisfiability problem ($K=2$ or $K=3$).
  
  The satisfiability transition in a  random $2$-SAT formula
  occurs when the clauses-to-variables ratio $\alpha$
  approaches $\alpha_c(2)=1$. However, long range frustration among 
  unfrozen variable nodes builds up only when 
  $\alpha > \alpha_R(2) = 4.4588$. For the random $3$-SAT problem, 
  we find a long--range frustrated mean field solution when 
  $\alpha > \alpha_R(3) = 4.1897$. The long range frustration
  order parameter $R$ of this solution jumps from zero to
  a finite positive value at $\alpha_R(3)$, while the energy density
  increases only gradually from zero as a function of $\alpha$.
  The SAT--UNSAT transition point of this solution is lower
  than the value of $\alpha_c(3)=4.267$ obtained by the survey 
  propagation algorithm. Two possible reasons for this discrepancy
  are suggested.
  
  The zero--temperature phase diagram of the $\pm J$ Viana--Bray
  model is also determined, which is identical to that of the
  random $2$-SAT problem. The predicted phase transition between a
  non-frustrated and a long--range frustrated spin glass
  phase might also be observable in real materials at a finite
  temperature. 

  \noindent
  [published in New Journal of Physics {\bf 7} (2005) 123; 
  freely available at
  http:{/}{/}www.njp.org{/}]
\end{abstract}
\pacs{89.75.-k, 75.10.Nr,  02.10.Ox}

\maketitle

\clearpage
                                %
                                %

\section{Introduction}

An attempt was made in a recent paper \cite{ZhouH2004b} to study the
vertex cover problem from the viewpoint of long range frustration. 
The vertex cover problem, sometimes also referred to as the hard--core gas
condensation problem in the physics literature,  can be mapped to a
spin glass model with quenched randomness coming from the underlying
random vertex--vertex connection patterns.
In the present paper, we extend the basic idea and the
method of Ref.~\cite{ZhouH2004b} to spin glass systems with an
additional kind of quenched randomness due to the random couplings
among interacting vertices. We demonstrate our calculations by working
on a spin glass model of the random $K$-satisfiability ($K$-SAT) problem.
Our results will be compared with known algorithmic and analytical conclusions.

The $K$-SAT is at the root of computational
complexity \cite{GareyM1979}. A $K$-SAT formula involves $N$ Boolean variables 
and $M= \alpha N$ constraints or clauses, each of which is a disjunction of $K$ Boolean
variables or their negations. The parameter $\alpha$ is called the
clauses-to-variables ratio. A $K$-SAT formula is satisfiable (SAT) if there
exists at least one assignment of the Boolean variables 
(a solution) such that all clauses are
satisfied; otherwise it is unsatisfiable (UNSAT). For example, the $2$-SAT
formula
$$
(x_1 \lor \bar{x_2}) \land (x_1 \lor x_2)
\land (\bar{x_1}\lor x_3)\land (x_2\lor \bar{x_3})
$$ 
with $N=3$ and $M=4$ has a SAT solution of  $x_1=x_2=x_3={\tt TRUE}$. 
Given a $K$-SAT formula, the
objective is to judge whether it is satisfiable or not, and if it is,
to find a satisfying solution.
The satisfiability of a $2$-SAT formula can be determined
in a run time that scales linearly with the number $N$ of
Boolean variables \cite{AspvallB1979}. However, to determine the 
satisfiability of a $K$-SAT formula with $K \geq 3$ 
is NP-complete, with a computation time that
scales {\em exponentially} with $N$ in the worst case.
The reason for this qualitative difference in computational complexity
is still not completely clear.

When a $K$-SAT formula of $N$ variables and $M$ clauses is chosen randomly
from the ensemble of all such formulas, 
it was observed  \cite{CheesemanP1991,KirkpatrickS1994,CrawfordJ1996} that
its probability $p$ of being
satisfiable drops from $p\approx 1$ to $p\approx 0$ over a small range of the
parameter $\alpha$ around certain 
$\alpha_c$ (for $K=3$, $\alpha_c \approx 4.2$
\cite{KirkpatrickS1994,CrawfordJ1996,AchlioptasD2001}).
Furthermore, although the $K$-SAT problem ($K\geq 3$) in general 
is NP-complete,
the satisfiability of a randomly chosen formula can often be
easily determined by a heuristic algorithm, provided that its
parameter $\alpha$ is {\em not} at the
vicinity of the SAT-UNSAT transition point $\alpha_c$.
The really difficult instances of random $K$-SAT ($K\geq 3$) are 
located at the parameter regime of $\alpha \approx \alpha_c$ \cite{CheesemanP1991}.

The above mentioned threshold phenomena are reminiscent of
what is usually observed in a physical system around 
its phase transition or critical point. Concepts of 
spin glass physics, such as replica symmetry breaking
and proliferation of metastable states, were  applied to the random
$K$-SAT problem in some recent articles
\cite{MonassonR1996,MonassonR1997,MonassonR1999,MezardM2002,MezardM2002b}. 
For the random $3$-SAT problem, M{\'{e}}zard and co-authors 
\cite{MezardM2002,MezardM2002b} found that in the parameter regime
of $3.921 < \alpha < 4.267$, there exists a hard-SAT phase with positive
structural entropy (complexity). A random $3$-SAT formula in this
phase is satisfiable by exponentially many Boolean configurations, 
which are clustered
into different macroscopic domains. There are also exponentially
many metastable domains of configurations which satisfy most but not
all the clauses (a local search algorithm is likely to be trapped in
one of these metastable configurational domains). 
This physical insight is implemented
in an efficient  heuristic 
survey propagation algorithm \cite{BraunsteinA2002}
to find a solution for a random $3$-SAT formula.

Trying to understand more deeply the qualitative
difference between the $2$-SAT and the general
$K$-SAT problem ($K \geq 3$), 
in this paper we re-examine the random $2$-SAT and 
$3$-SAT problem by focusing on the issue of long range frustration 
among unfrozen
Boolean variables.
A long range frustration order parameter $R$ is calculated.
We find that, while the SAT--UNSAT transition occurs at 
$\alpha_c(2)=1$ for the random $2$-SAT problem,
long range frustration only gradually builds up when 
$\alpha \geq \alpha_R(2) = 4.4588$, where
the non-frustrated mean field solution becomes unstable.
For the random $3$-SAT problem, we find a self-consistent
long--range frustrated solution when $\alpha \geq \alpha_R(3)= 4.189724$.
The long range frustration
order parameter $R$ of this solution jumps from zero to
a finite positive value at $\alpha_R(3)$, while the energy density
increases only gradually from zero as a function of $\alpha$.
This solution predicts that the SAT-UNSAT transition occurs
at $\alpha = \alpha_R(3)$.
However, for the random $3$-SAT problem, the non-frustrated
trivial mean field solution is stable even when $\alpha > \alpha_R(3)$;
and furthermore, in the range of $3.921 < \alpha < 4.267$ there is
a non-frustrated replica symmetry breaking solution of
zero energy \cite{MezardM2002b}. It is likely that the
true SAT-UNSAT transition point is at $\alpha = \alpha_c(3) = 4.267$.
Two possible reasons for this discrepancy between our mean field
solution and the solution of Ref.~\cite{MezardM2002b} 
are discussed in this paper. 
It appears that, when the replica symmetry breaking occurs
before the onset of long range frustration, the method outlined
in this paper needs to be improved, and competitions
among different macroscopic states should be considered
in the mean field theory.
More work is certainly needed to overcome this discrepancy.

This paper is organized as follows.
We first define the random $K$-SAT
problem in Sec.~\ref{sec:model}. The random $2$-SAT problem 
and the $\pm J$ Viana--Bray spin glass model are
then studied in Sec.~\ref{sec:2}.  The method developed in
this section is then applied to the more difficult
random $3$-SAT problem in Sec.~\ref{sec:3}.
We summarize and discuss our main results in Sec.~\ref{sec:4}.

                                %
                                %

\section{
  Spin glass model of the random $K$-satisfiability problem
  \label{sec:model}
}

We follow Ref.~\cite{MezardM2002} and represent a random $K$-SAT formula
by a random factor graph
of $N$ variable nodes and $M=\alpha N$ 
function nodes. Each variable node $i$ represents a Boolean variable,
with a spin value $\sigma_i= \pm 1$.  It is connected to $k$ function nodes,
where $k$ is a random integer governed by the Poisson distribution 
$f(k, c)$ of mean $c = K \alpha$. The Poisson distribution is defined as
\begin{equation}
  \label{eq:poisson}
  f( k , c ) ={ e^{-c} c^k  \over k !} \ .
\end{equation}
Each function node $a$ represents
a clause (constraint); it is linked to $K$ randomly chosen 
variable nodes  $i_1,\ldots, i_K$. Function node $a$ has an 
energy $E_a$ which is either $0$ (clause satisfied) or $1$ (clause
violated):
\begin{equation}
  E_a= \prod\limits_{r=1}^K {1-J_{a}^{i_r} \sigma_{i_r} \over 2} \  ,
  \label{eq:01}
\end{equation}
where $J_{a}^{i_r}$ is the edge strength between
$a$ and $i_r$:
$J_{a}^{i_r}=-1$ or $1$ depending on whether 
Boolean variable $i_r$ in clause $a$ is negated or not. 
In a given  random $K$-SAT formula, $J_{a}^{i_r}$ is  a 
quenched random variable with value equally
distributed over  $\pm  1$.  The total energy of the system for 
each of the $2^N$ possible spin configurations is
expressed as 
\begin{equation}
  \label{eq:totalenergy}
  E[\{ \sigma_i  \}]=\sum\limits_{a=1}^{M} E_a \ .
\end{equation}
As we are interested in the ground--state energy landscape of
model Eq.~(\ref{eq:totalenergy}), only 
those configurations which have the global minimum energy are
considered in later discussions.

                                %
                                %

\section{
  The random $2$-satisfiability problem and the $\pm J$ 
  Viana--Bray spin glass  model
  \label{sec:2}
}

We begin with the random $2$-SAT problem. Experience gained 
in this section 
will help us to understand the more difficult random $3$-SAT
problem in the next section.
We discuss the random $2$-SAT problem first 
from the viewpoint of long range frustration
(Sec.~\ref{sec:2a}). Then we study it in
Sec.~\ref{sec:2b} through the cavity method of M{\'{e}}zard and
Parisi \cite{MezardM2001,MezardM2003} by setting the
long range frustration order parameter to $R=0$.
The later treatment
causes  an annoying divergence in the population dynamics, suggesting
the necessity of taking into account the effect of
long range frustration.
In Sec.~\ref{sec:2c} we briefly discuss the 
finite connectivity $\pm  J$ Viana--Bray spin glass model.

\subsection{
  Long range frustration in a single macroscopic state
  \label{sec:2a}
}

The ground--state configurations of a random $2$-SAT formula
may be classified into different macroscopic states 
\cite{MezardM2003} (hereafter, a macroscopic state is
simply referred to as a state). 
Two microscopic configurations in the same state are mutually reachable by 
flipping a  finite number of spins 
of one configuration and then 
letting the system relax. According to this definition of
states, two configurations of the same state
may differ from each other in the spin values of $O(N)$ variable
nodes (this is due to the formation of a percolation cluster
of unfrozen variable nodes discussed later). 
We focus on a randomly chosen macroscopic
state $\beta$. In state $\beta$,  the spin value of a randomly chosen 
variable node $i$ may be
fixed to $\sigma_i \equiv +1$ (positively frozen), or to $\sigma_i \equiv -1$
(negatively frozen), or fluctuate over  $\pm 1$ (unfrozen).  
The fraction of positively frozen, negatively
frozen, and unfrozen variable nodes in state $\beta$ is, respectively, 
$q_{+}$, $q_{-}$, and $q_{0}$, with $q_{+} + q_{-} + q_{0} = 1$. 
Our approach at the present stage is essentially replica symmetric
because of the following reasons:
(a) although many macroscopic states may coexist, our system is
assumed to stay always in the same state,(b) a state is
characterized by just two mean field parameters, $q_0$ and $R$
(to be defined later).

Since the spin values of the unfrozen variable nodes fluctuate among
different configurations of state $\beta$, the possibility of correlation
among these fluctuations must be
taken into account.  Consider an unfrozen variable node $i$. 
If its spin value is externally
fixed to a prescribed value $\sigma_i = \sigma_i^*$ ($\sigma_i^*=\pm
1$ with equal probability),
how many other unfrozen variable nodes must
eventually fix their spins as a consequence?

For a random factor graph with size $N \to \infty $, 
the total number $s$ of affected variable nodes may scale linearly
with $N$ and therefore also approaches infinity.
If this happens, $\sigma_i^*$ is referred to as a canalizing value for
variable node $i$, and variable node $i$ is referred to as
type-I unfrozen with respect to spin value $\sigma_i^*$. 
This happens with probability
$R$, with $R$ being our long range frustration order parameter.
The total number of affected variable nodes may also be finite,
$s\sim O(1)$. 
In this case, variable node $i$ is  type-II unfrozen
with respect to spin value $\sigma_i^*$.  
We emphasize that, if an unfrozen
variable node $i$ is type-I unfrozen with respect to spin
value say $\sigma_i=+1$, it is not necessarily also
type-I unfrozen with respect to spin value $\sigma_i=-1$;
in other words, the fact that $\sigma_i^*$ is a canalyzing value for
variable node $i$ does not mean that $-\sigma_i^*$ is also
a canalyzing value for the same node. In the remaining part of
this article, to simplify the notation,
we will frequently use phrases such as
$``$variable node $i$ is type-I unfrozen'' and
$``$a type-II unfrozen variable node $i$''. 
It should be understood that, whenever we mention that
a variable node $i$ is type-I or type-II unfrozen, we
mean that it is type-I or type-II unfrozen 
with respect to a prescribed spin value $\sigma_i^*$. 

Based on the bond percolation theory on random graphs 
\cite{BollobasB1985}, we know that the
percolation clusters evoked by two type-I unfrozen variable nodes 
have a nonzero intersection of size proportional to $N$.
Therefore, the spin values of all type-I unfrozen variable nodes
must be strongly correlated. If we randomly choose two
type-I unfrozen variable nodes $i$ and $j$,
then with probability {\em one half}
$\sigma_i^*$ and $\sigma_j^*$  can not be realized simultaneously
in any configuration of state $\beta$:  
if $\sigma_i=\sigma_i^*$, then $\sigma_j$ must take 
the value $\sigma_j=-\sigma_j^*$; 
if $\sigma_j=\sigma_j^*$, then $\sigma_i$ must take the value 
$\sigma_i=-\sigma_i^*$. 
The probability one half comes from the fact that both $\sigma_i^*$ 
and $\sigma_j^*$ are
random variables equally distributed over  $\pm  1$ (due to the quenched
randomness in the edge strengths, as will become clear later).
On the other hand, two randomly chosen type-II unfrozen 
variable nodes are mutually independent, since each of them can only 
influence the spin values of $s \sim O(1)$ other variable 
nodes while the shortest path length between them scales as $\ln N$
\cite{BollobasB1985}. 
For later convenience, we denote $F(s)$ as the probability that a
randomly chosen unfrozen
variable node $i$ will eventually fix a total number $s$ of
other unfrozen variable nodes when it is flipped to a prescribed value 
$\sigma_i= \sigma_i^*$,
where $s$ is under the constraint that $s\sim O(1)$ and
therefore $\lim_{N\to \infty} s/N=0$. 

We follow the cavity field approach to calculate the parameters $q_{+}$, $q_{-}$, 
and $q_{0}$.  We first generate a random factor graph $G$ of $N$ variable nodes and
(on average) $\alpha N$ function nodes.
Then we create a new variable node $i$ and connect it to 
$G$ by a set $V_i$ of new function nodes, the size $k$ of this set
following the Poisson distribution $f(k, 2 \alpha)$.
Each of these newly constructed function nodes is connected
to a randomly chosen variable node of $G$ at its other end.
The enlarged factor graph is denoted as $G^\prime$. 
It has $N+1$ variable nodes and on average
$\alpha N + 2 \alpha$ function nodes.

We look at the local environment of a newly added
function node $a$, which is connected to the new variable node $i$ and
a randomly chosen variable node $j$ of factor graph $G$.
Let us denote $\sigma_j^* = J_a^j$. 
In factor graph $G$ variable node $j$ may be frozen to the spin value  $\sigma_j^*$ or
to $-\sigma_j^*$, it may also be unfrozen.
Now we ask the question:  Can function node $a$  be satisfied by variable
node $j$ alone? There are the following possibilities:

\begin{description}
\item[(i).]
  Node $j$ is frozen in factor graph $G$ to $\sigma_j\equiv \sigma_j^*$.
  This happens with probability $(q_{+} + q_{-})/2 = ( 1- q_0 ) / 2$. 
  In this situation, function node $a$ is satisfied.
  
\item[(ii).]
  Node $j$ is unfrozen and $\sigma_j^*$ is not a canalizing value of
  node $j$ (type-II unfrozen).
  This happens with probability $ q_0 ( 1 - R)$. 
  In this situation, function node $a$ is also satisfiable 
  by flipping $\sigma_j$ to the value $\sigma_j^*$.
  
\item[(iii).]
  Node $j$ is unfrozen and $\sigma_j^*$ is a canalizing value of node
  $j$ (type-I unfrozen). 
  This happens with probability $ q_0 R$. 
  In this situation,
  function node $a$ is  also satisfiable by flipping $\sigma_j$ to 
  the value $\sigma_j^*$.  
  However,  this spin flip will evoke a  percolation cluster 
  (with size scaling linearly with $N$) of unfrozen variable nodes,
  namely, the spin values of these unfrozen variable nodes 
  all have to be fixed.
  
\item[(iv).]
  Node $j$ is frozen to $\sigma_j=-\sigma_j^*$.
  This happens with probability $(q_{+} + q_{-})/2 = ( 1 - q_0 ) / 2 $.
  In this situation, function node $a$ is unsatisfiable by node $j$. 
  To  satisfy $a$, variable node $i$ must be fixed to the 
  value $\sigma_i = J_{a}^i$.
\end{description}

In the set $V_i$, a number $k_{iii}^{+}$  of function nodes,
each connected to variable node $i$ by a positive bond,
are facing the local environment described by the above--mentioned
situation (iii).
The random integer $k_{iii}^{+}$ is governed by the Poisson
distribution $f( k_{iii}^{+}, \lambda_1 )$, where
$\lambda_1 = \alpha q_0 R$. When $\sigma_i$ is set to
$\sigma_i=-1$, these function nodes are not satisfied by node $i$. To
satisfy them, the $k_{iii}^{+}$ 
unfrozen variable nodes attached to the other ends of 
these function nodes must all be flipped to the appropriate spin values. 
However, since each such spin flips will lead to a
percolation cluster of size approaching infinity, 
these function nodes may not all be 
simultaneously satisfiable 
due to the possibility of long range frustration among
the type-I unfrozen variable nodes. 
These $k_{iii}^{+}$ type-I unfrozen variable nodes can be divided into
two subgroups (say $S_A$ and $S_B$); variable nodes of the same subgroup can take 
their canalyzing
spin values simultaneously in state $\beta$, while variable nodes from
different subgroups will never take their corresponding
canalyzing spin values simultaneously in state $\beta$. 
Because all these $k_{iii}^{+}$ variable nodes are
unfrozen, their spin values for sure will fluctuate among different
miscroscopic configurations. Therefore, in some microscopic configurations,
variable nodes in $S_A$ will take their canalyzing spin values; and in
some other microscopic configruations, variable nodes in $S_B$ will
take their canalyzing spin values.
When the spin value of variable node $i$ is set to $\sigma_i=-1$, 
the probability $P_{\rm f}(n)$ that at least $n$ function nodes will
be unsatisfied due to this type of long range frustration can be expressed as
\begin{equation}
  P_{\rm f}(n) =  f( 2 n , \lambda_1 ) \ C_{2n}^{n}\  2^{- 2 n} + 
  \sum\limits_{n^\prime=2 n +1}^{\infty} f(n^\prime,  \lambda_1 )
  \ C_{n^\prime}^{n} \  2^{1-n^\prime} \ , 
  \label{eq:05}
\end{equation}
where $C_{n^\prime}^{n}=n^\prime ! / ( n ! (n^\prime -n)!)$ is the binomial
coefficient. 
The first term on the right hand side of 
Eq.~(\ref{eq:05}) corresponds to the case that 
the above-mentioned two subgroups have the
same size $n$ ($k_{iii}^{+}= 2n$); the second term  corresponds to
the case that these two subgroups are not equal in size,
one containing $n$ variable nodes and the other containing
more than $n$ variable nodes ($k_{iii}^{+} \geq 2 n +1$).

In the set $V_i$ of function nodes, $k_{iv}^{+}$ of them are
connected to $i$ by a positive bond and are facing the local
environment (iv). The random value $k_{iv}^{+}$ is governed 
by the Poisson distribution
$f( k_{iv}^{+},  \lambda_2 )$, with
$\lambda_2 = \alpha ( 1 - q_0 )/2$.  These function nodes are
definitely unsatisfied when $\sigma_i = -1$. 
Therefore, the probability $P_{\rm v}(m_{-})$
that $m_{-}$ function nodes will be unsatisfied when $\sigma_i = -1$ is
\begin{equation}
  \label{eq:04}
  P_{\rm v}(m_{-})=\sum\limits_{n=0}^{m_{-}} 
  f\bigl( n , \lambda_2  \bigr) \  P_{\rm f}(m_{-} - n) \ .
\end{equation}

When $\sigma_i$ is set to $\sigma_i = +1$, $m_{+}$ function nodes 
in the set $V_i$ may be unsatisfied. 
It is easy to verify that
$m_{+}$ follows the same probability distribution as $m_{-}$.
Since the probability distributions for $m_{+}$ and $m_{-}$ are
known, the probability $q_0(i)$ for variable node $i$ to be
unfrozen in the corresponding state $\beta^\prime$ of 
the enlarged graph $G^\prime$ can be 
calculated. In the large $N$ limit, we assume the following
convergence condition that 
\begin{equation}
  \lim\limits_{N\to \infty} q_0(i) = q_0 \ .
  \label{limit}
\end{equation}
With this assumption, we can therefore
write down the following self-consistent equation
for  the parameter $q_0$:
\begin{equation}
  q_0 = \sum\limits_{m_{+}=0}^{\infty} \sum\limits_{m_{-}=0}^\infty
  P_{\rm v} (m_{+})\  P_{\rm v}(m_{-}) \ \delta_{m_{+}}^{m_{-}} \ ,
  \label{eq:06}
\end{equation}  
where $\delta$ is the Kronecker symbol. 
It is easy to verify that $q_{+} = q_{-} = (1 - q_0 ) / 2$.

We make an important remark here. At the present stage of the
mean field theory, we focus only on one state $\beta$ of
factor graph $G$, and assume
that Eq.~(\ref{limit}) holds in this state $\beta$.
When the system has more than one state, there may be
competitions among these different states. As an example
of such competitions, consider two
states $\beta_1$ and $\beta_2$ of factor graph $G$.
We assume $\beta_1$ is a ground state and $\beta_2$ is a ground state or
metastable state, therefore $E_1 \leq E_2$, where $E_{1}$ 
($E_2$) is
the energy of factor graph $G$ in state $\beta_1$ $(\beta_2)$.
When the new variable node $i$ is added,
$n_1$ and $n_2$ of the nearest-neighbor function nodes of
$i$ may be violated in the two corresponding states $\beta_1^\prime$
and $\beta_2^\prime$ of  
factor graph $G^\prime$. Therefore, $G^\prime$  have
energy $E_1^\prime= E_1 + n_1$ and $E_2^\prime=E_2 + n_2$, respectively,
in state $\beta_1^\prime$ and $\beta_2^\prime$. We see that, although 
$E_1 \leq E_2$, in the enlarged system state $\beta_2^\prime$  may
have lower energy than state $\beta_1^\prime$ if 
$n_1 - n_2 > E_2 - E_1$. To correctly predict the
ground state energy of the enlarged system $G^\prime$, we
may need the information on all the low energy states of
the original system $G$. We hope to return to this
point in a future work. In the present paper, we stick
to one state $\beta$ of factor graph $G$.

Now we calculate the order parameter $R$. Suppose in state $\beta$
variable node
$i$ is unfrozen, and it is type-II unfrozen with respect
to the spin value, say, $\sigma_i = -1$. That is, flipping the spin value of
node $i$ to $\sigma_i = -1$ will cause the fixation of 
the spin values
of $s\sim O(1)$ other unfrozen variable nodes. Then all the function nodes
that are connected to $i$ by a positive bond should not face the
above--mentioned local environment  (iii) but may face the local
environment (ii). Consider such a function node $a$, and suppose
its other end is connected to a type-II unfrozen variable node $j$
by  a positive bond (the discussion is the same if the bond is
negative).
When the spin value of variable node $i$ is set to $\sigma_i = -1$, 
the spin value of node $j$ must be 
flipped to $\sigma_j = +1$
to satisfy function node $a$. 
On the other hand, variable node $j$
may be connected to other function nodes besides node $a$; and
therefore when $\sigma_j$ is set to $\sigma_j=+1$, some other 
type-II unfrozen variable nodes will be affected, $\ldots$ 
(a $``$chain reaction").
Based on  this observation,
the probability distribution $F(s)$ defined earlier in this subsection
can be determined by the following iterative equation:
\begin{equation}
  F(s)  =  f\bigl(0,  \lambda_3 \bigr) \  \delta_{s}^{0} \ + 
  \sum\limits_{k=1}^{\infty} f\bigl( k , \lambda_3 \bigr)
  \sum\limits_{\{s_l \}}
  \delta_{s_1+\ldots +s_k }^{s} \prod\limits_{l=1}^{k} F(s_l)\ ,
  \label{eq:02}
\end{equation}
where $\lambda_3 = \alpha q_0 ( 1 - R )$ is the average 
number of function nodes that are attached to variable node
$i$ by a positive bond and are facing the above mentioned
local environment (ii).
Equation~(\ref{eq:02}) holds true as long as the cluster of
affected variable nodes forms a tree, i.e., there is no loops. 
This constraint is satisfied in a random factor graph
of size $N\to \infty$, since the shortest loop
length between two variable nodes scales as $\ln N$ 
\cite{BollobasB1985} while the cluster size $s$ scales
as $O(1)$.

Since $R=1-\sum_{s=0}^\infty F(s)$, from Eq.~(\ref{eq:02}) we realize that
\begin{equation}
  R=1- \exp\bigl(-  \lambda_3 R \bigr) \ .
  \label{eq:03}
\end{equation}

\begin{figure}[t]
  \psfig{file=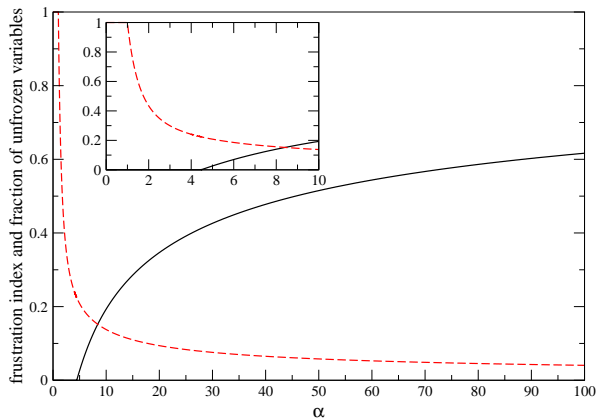,angle=270,width=10.0cm}
  \caption{ \label{fig:02}
    The long range frustration order parameter $R$ (solid lines) and
    the unfrozen probability $q_0$ (dashed lines) as a function 
    of $\alpha$ for the
    random $2$-SAT problem.   
    The inset is an amplification of the region of
    $ \alpha \in [0 , 10]$.
  }
\end{figure}

The unfrozen probability $q_0$ and the long range frustration order
parameter $R$, as calculated by Eqs.~(\ref{eq:06}) and (\ref{eq:03}),
are shown in Fig.~\ref{fig:02} as a function of
the clauses-to-variables ratio $\alpha$.  Onset of spin freeze occurs at
$\alpha = \alpha_c(2) = 1$. This point, as shown later, is also the point of
SAT--UNSAT transition, in agreement with the
rigorous mathematical proof \cite{FernandezW2001}. 
A typical random $2$-SAT formula of $\alpha < 1$ is
satisfiable while a typical random $2$-SAT formula of $\alpha > 1$ is
unsatisfiable. However, onset of long range frustration is delayed 
to $\alpha = \alpha_R(2) = 4.4588$. 
For $\alpha_c(2) < \alpha < \alpha_R(2)$ a typical random
$2$-SAT formula, although being unsatisfiable, is not long--range 
frustrated. 
We know that to determine the satisfiability of a $2$-SAT formula is
computationally easy. It is interesting to
point out that,  long range
frustration is absent at {\em both} sides of 
the SAT-UNSAT transition $\alpha_c(2)$ for the random $2$-SAT problem.

Based on our calculation that a randomly constructed $2$-SAT formula
in the parameter regime of $\alpha_c(2) < \alpha < \alpha_R(2)$ is
not long--range frustrated ($R=0$), 
we hope that a Boolean assignment that
satisfies the largest possible number of its clauses
might be easily found in practice, despite the fact
that the MAX-$2$-SAT problem is in general
NP-hard \cite{FernandezW2001}.
We will check this point by computer 
simulations in a later work.

When $\alpha \geq \alpha_R(2)$, long range frustrations exist among a 
fraction of
the unfrozen variable nodes. Interestingly, the point 
$\alpha= \alpha_R(2)$ also marks the
beginning of divergence of the population dynamics of the next
subsection.

Now we turn to the ground--state energy density $\epsilon$. The 
energy increase $\epsilon_{1}$ due to the addition of variable node $i$ is
\begin{equation}
  \epsilon_1 (\alpha) = \sum\limits_{m_{+}=0}^\infty 
  \sum\limits_{m_{-}=0}^\infty P_{\rm v}(m_{+}) \
  P_{\rm v}(m_{-}) \  \min( m_{+}, m_{-} ) \ .
  \label{eq:07}
\end{equation}
After the addition of variable node $i$, the factor graph $G^\prime$
has $N+1$ variable nodes and on average $\alpha N + K \alpha$ function nodes
($K = 2$).
Its clauses-to-variables ratio is therefore
$\alpha^\prime = \alpha + (K-1) \alpha / (N+1)$. Suppose the energy 
density $\epsilon$ of the system is
a function only of $\alpha$. Then from the identity 
$$
(N+1) \epsilon( \alpha^\prime )  = N \epsilon( \alpha ) 
+ \epsilon_1 \ ,
$$
we obtain the following differential equation
$$
\epsilon( \alpha ) + ( K - 1 ) \alpha  
{ {\rm d} \epsilon \over {\rm d} \alpha } = \epsilon_1 
$$
in the limit of large $N$. By solving this differential
equation, we obtain that
\begin{equation}
  \epsilon(\alpha)= { 1 \over ( K-1) \alpha^{1/(K-1)} }
   \int_{0}^{\alpha} \tilde{\alpha}^{{2-K \over K-1 } } 
   \epsilon_1( \tilde{\alpha} ) {\rm d} \tilde{\alpha} \ ,
  \label{eq:08}
\end{equation}
where $K=2$.
According to the author's unpublished calculations,
in the vertex cover problem, the energy density derived through this
way is the exact ground--state energy density when there is
no long range frustration; however, when $R > 0$, 
this energy density is  lower than the
actual ground--state energy density obtained
by Weigt and Hartmann by numerical means \cite{WeigtM2000} 
(the reason, however, is still 
unknown). We expect the same situation will
occur here in the random $K$-SAT problem.

When long range frustration exists ($R > 0$), an upper bound for 
the ground--state energy density $\epsilon$ can also be derived.
For the enlarged factor graph $G^\prime$ to have clauses-to-variables 
ratio  $\alpha$,
on average $(K-1) \alpha$ function nodes must be removed
\cite{MezardM2003}. The energetic contribution 
of these function nodes must also be removed. This contribution
comes from two parts: (i) the energy sum of these individual function nodes,
$\Delta E= (K-1) \alpha [(1-q_0)/2]^K$ and, (ii)
additional energy $\Delta E^\prime$ caused by  possible
long range frustration among  these function nodes. Therefore,
\begin{equation}
  \epsilon( \alpha ) = \epsilon_1 - \Delta E - \Delta E^\prime.
  \label{eq:09}
\end{equation}
If we set $\Delta E^\prime =0$, Eq.~(\ref{eq:09}) then gives an upper bound.

\begin{figure}[t]
  \psfig{file=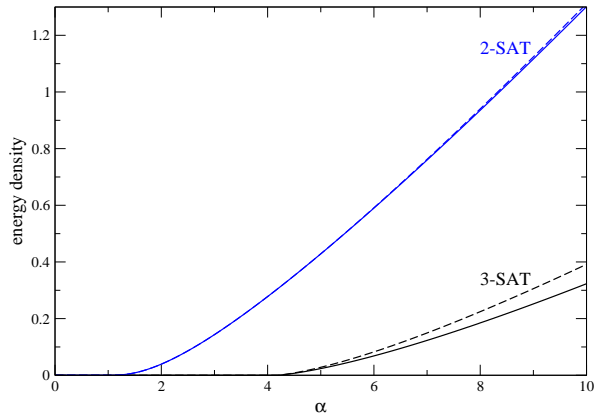,angle=270,width=10.0cm}
  \caption{ \label{fig:03}
    The energy density lower bound (solid lines) and 
    upper bound (dashed lines) for the random $2$-SAT and
    $3$-SAT problem as a function of $\alpha$.
  }
\end{figure}

The energy densities calculated based on Eqs.~(\ref{eq:08}) and (\ref{eq:09}) 
are shown in Fig.~\ref{fig:03}. When $\alpha > \alpha_c(2)$, the energy density
becomes positive and the system is UNSAT.  In the range of 
$\alpha_c(2) \leq \alpha < \alpha_R(2)$,
both Eq.~(\ref{eq:08}) and (\ref{eq:09}) give the identical results,
indicating that when there is no long range frustration, these equations
are exact. When $\alpha \geq \alpha_R(2)$, the energy density given 
by Eq.~(\ref{eq:09}) is
systematically higher than that given by Eq.~(\ref{eq:08}). In this regime of
long range frustration,   at the moment 
the author is still unable to give an exact
ground--state energy density expression. We hope to return to this point
in a future work.

\subsection{
  Random $2$-satisfiability studied by population dynamics
  \label{sec:2b}
}

In this subsection, we study the random $2$-SAT problem 
following the zero--temperature cavity method treatment
of Ref.~\cite{MezardM2002b}. Since in the parameter
range of $0 \leq \alpha < \alpha_R(2)$ there is
no long range frustration, we expect this treatment to
be exact in this range. 

Suppose there are many macroscopic states of global
minimum  energy.
Let us randomly choose a function node $a$ and 
follow one of its edges to a variable node $i$. Before the
edge $(a, i)$ is connected, variable node
$i$ may be positively frozen in some macroscopic states,
and be negatively frozen in some other macroscopic states,
and be unfrozen in the remaining macroscopic states. 
A positively (negatively) frozen variable node is said to feel a 
positive (negative) cavity field $h_i$, while an unfrozen 
variable node is said to feel a zero cavity 
field $h_i$ \cite{MezardM2003}. 
The (modified) cavity field $J_a^i h_i$ distribution among all 
the macroscopic states
(referred to as the cavity field survey)
is denoted by $P( J_a^i h_i )$, where $J_a^i$ is the
bond strength between function node $a$ and variable node $i$.
Besides function node $a$, node  $i$ is connected by
positive bonds to $p$ other function nodes $b_1 , \ldots , b_p$ and
by negative bonds to $n$ other function nodes $c_1 , \ldots , c_n$.

Assuming there is no long range frustration among the unfrozen
variable nodes ($R=0$), one can write down the following self-consistent
equation for the cavity field distribution $P(J_a^i h_i)$
\cite{BraunsteinA2002}:
\begin{eqnarray}
  P(J_a^i h_i) & 
  \propto & \int \bigl[ \prod\limits_{r=1}^{p} P( J_{b_r}^{j_r} h_{j_r} )
  {\rm d} h_{j_r} \bigr]
  \bigl[ \prod\limits_{l=1}^{n} P( J_{c_l}^{k_l} h_{k_l} )
  {\rm d} h_{k_l} \bigr] \times \nonumber \\
  & & 
  \delta\bigl( 
  J_a^i h_i - \sum\limits_{r=1}^{p} \theta(-J_{b_r}^{j_r} h_{j_r} 
  \bigr)
  + \sum\limits_{l=1}^n \theta(-J_{c_l}^{k_l} h_{k_l} ) ) \times \nonumber \\
  & & \exp(-y \Delta E_1 /2 )
  \label{eq:survey}
\end{eqnarray}
where $\theta(x)$ is the Heaviside step function, $\theta(x)=1$ if
$x >0$ and $\theta(x)=0$ otherwise; $y$ is a re-weighting parameter;
and 
\begin{eqnarray}
  \Delta E_1 &=&
  \sum_{r=1}^p \theta(-J_{a_r}^{j_r} h_{j_r} )
  + \sum_{l=1}^n \theta(-J_{b_l}^{k_l} h_{k_l}) \nonumber \\
  & & 
  - |\sum_{r=1}^p \theta(-J_{a_r}^{j_r} h_{j_r} ) 
  - \sum_{l=1}^n \theta(-J_{b_l}^{k_l} h_{k_l}) |
  \label{eq:d_E_1}
\end{eqnarray}
is twice the energy caused by the function nodes $b_1,\ldots, b_p,
c_1,\ldots, c_n$.

In the limit of $y\to +\infty$,
the general form of the cavity field distribution is \cite{BraunsteinA2002}
\begin{equation}
  P( J_a^i h_i ) =
  \left\{
    \begin{array}{ll}
      P_{+}( J_a^i h_i)\ , &  p_+ \\
      P_{-}( J_a^i h_i)\ , &  p_- \\
      \begin{array}{r}
        \eta_0(i) \delta( J_a^i h_i ) + \eta_+(i) P_{+}( J_a^i h_i )
        \\
        + \eta_-(i) P_{-}( J_a^i h_i )\ .
      \end{array} & p_0 
    \end{array}
  \right.
  \label{eq:survey_form}
\end{equation}
In Eq.~(\ref{eq:survey_form}), $p_{+}$, $p_{-}$, and $p_0$ are the
probabilities of occurrence for the three types of cavity field
surveys. $P_{+}(k)$ is a probability distribution over 
positive integers $k$; and $P_{-}(k)$ is a probability distribution
over  negative integers $k$. The parameters $\eta_0(i)$,
$\eta_{+}(i)$, and $\eta_{-}(i)$ are three random variables,
$\eta_{0}(i)+\eta_{+}(i)+\eta_{-}(i)=1$.

Based on Eqs.~(\ref{eq:survey}) and (\ref{eq:survey_form}), 
one finds that
\begin{eqnarray}
  p_0 = \sum\limits_{m=0}^\infty
  {  e^{- \alpha ( 1 - p_0 ) } \bigl( \alpha ( 1 - p_0 ) / 2 \bigr)^{2 m}
    \over ( m! )^2 } \ ,
  \label{eq:p_0} \\
  p_+ = p_- = ( 1 - p_0 ) / 2 \ .
  \label{eq:p+}
\end{eqnarray}

When $\alpha \leq  \alpha_R(2)$, the probability $p_0$ satisfies
$\alpha p_0 \leq 1$.  Consequently,
the population dynamics at $y=\infty$ converges to the 
replica symmetric fix point of $\eta_{0}\equiv 1, \eta_{+} =\eta_{-}\equiv 0$. 
In other words, the system has only one macroscopic state. 
When $\alpha > \alpha_R(2)$, this replica symmetric fixed point
is unstable, and if initially the random variables $\eta_{0}(i)$ are
slightly less than unity, 
the population dynamics at $y=\infty$ becomes divergent.  
This divergence suggests that, at this parameter regime,
the assumption of absence of long range frustration becomes
invalid. 

To get rid of this divergence, 
one can run population dynamics at a finite positive value 
of $y$. A conceptually 
better way is to directly take into account the
effect of long range frustration in the population dynamics.
Unfortunately, at the moment the author is
unable to achieve this goal.

\subsection{
  The $\pm J$ Viana--Bray spin glass model
  \label{sec:2c}
}

We now briefly discuss the zero temperature phase diagram of the
$\pm J$ Viana--Bray model \cite{VianaL1985}. The model is characterized
by the Hamiltonian
\begin{equation}
  \label{eq:10}
  E[\{ \sigma_i \}]= - \sum\limits_{(ij)} J_{i j} \sigma_i \sigma_j \ ,
\end{equation}
where the summation is over all the edges $(ij)$ of a Poisson random
graph of mean vertex degree $c$;  the edge strength $J_{ij}=\pm 1$,
with equal probability;
and $\sigma_i=\pm 1$ is the spin value on vertex $i$. The zero temperature
phase diagram of this model was first reported in Ref.~\cite{KanterI1987}
through the replica approach.
It was found that, when $c\leq  1$, the system is in the paramagnetic 
phase with no frozen vertices, i.e., $q_0 = 1$; when $c >  1$, 
the system is in the spin glass phase, where some vertices
are frozen to positive or negative
spin values, and $q_0$ gradually decreases from unity. 

The Hamiltonian Eq.~(\ref{eq:10}) is similar to that of the random $2$-SAT and
can be studied by the same method mentioned in subsection \ref{sec:2a}.
We find that
the zero temperature phase diagram of model (\ref{eq:10}) is
identical to Fig.~\ref{fig:02} of the random $2$-SAT. 
The transition between the
paramagnetic phase and the spin glass phase occurs at $c=1$, confirming the
earlier result of Ref.~\cite{KanterI1987}.  
(When the long range frustration order parameter is set to
$R=0$, the order parameter $q_0$  has the same expression
as Eq.~(15) of Ref.~\cite{KanterI1987}.) 
In the spin glass phase, long range frustration among unfrozen
vertices only builds up ($R > 0$) when the mean vertex 
degree $c \geq 4.4588$. 

The $\pm J$ finite connectivity spin glass model therefore has two types of
phase transitions as the mean connectivity is increased. First there is
the frozen transition to the spin glass phase; and then there is the 
transition to a long--range frustrated spin glass phase. 
It is more experimentally relevant to discuss the phase transitions in
terms of temperature. It is accepted that the spin glass phase will
persist over a finite temperature range of $0 \leq T < T_g$
\cite{VianaL1985}.
We expect that when the mean vertex degree $c > 4.4588$,
the long--range frustrated spin glass phase
will also persist over a temperature range of $0 \leq T < T_R$, with
$T_R \leq T_g$.  It will be of great interest to check the validity of
this picture by computer simulations.

Real world spin glass materials are three--dimensional, and their
spin--spin interactions may be more complex than that given by
Eq.~(\ref{eq:10}).
It is not known whether or not long range frustrations exist among the
unfrozen vertices of such systems, and if they do exist, how to
detect them experimentally. It may be useful to re-interprete 
the already documented 
experimental observations in terms of the long range frustration
order parameter $R$.

                                %
                                %

\section{
  The random $3$-satisfiability problem
  \label{sec:3}
}

This section focuses on the random $3$-SAT problem. 
We study it using the method of Sec.~\ref{sec:2a} and then
compare the results with those obtained by population dynamics.

The probability of a randomly chosen variable node being
positively frozen, negatively frozen, or unfrozen in a
macroscopic state $\beta$ of global minimum energy is 
denoted respectively as $q_{+}$, $q_{-}$, and $q_{0}$
as in the case of the random $2$-SAT problem. $R$ is the
probability that an unfrozen variable node is type-I 
unfrozen with respect to a prescribed spin value.
The parameters $q_0$, $q_+$, $q_{-}$, and $R$ 
can be obtained following the same procedure as was outlined in 
Sec.~\ref{sec:2a}.
We first generate a factor graph $G$ of $N$ variable nodes and
$M= \alpha N$ function nodes; we then create a new variable node $i$
and connect this new node $i$ to $G$ through the formation of
$k$ new function nodes, with $k$ following the Poisson distribution
$f(k, 3 \alpha)$. The other ends of such a function node $a$
are connected to two randomly chosen variable nodes $j$ and $k$
of the factor graph $G$. 
For convenience of discussion, we denote $\sigma_j^* = J_a^j$ and
$\sigma_k^* = J_a^k$; and in the remaining part of this section,
when we mention that a variable node $j$ is type-I or type-II
unfrozen, we mean that it is type-I or type-II unfrozen with respect
to the spin value $\sigma_j^*$. In state $\beta$,
the function node $a$  faces one  of the
following local environments:
\begin{description}
\item[(i).]
  Node $j$ is frozen in factor graph $G$
  to $\sigma_j\equiv \sigma_j^*$ or 
  node $k$ is frozen to $\sigma_k= \sigma_k^*$. 
  This happens with probability $( 3 - 2 q_0 - q_0^2 ) / 4$.
  In this situation, function  node $a$ is satisfied by $j$ 
  or  $k$ (or both).
  
\item[(ii).]
  Both node $j$ and node $k$ are type-II unfrozen.
  This  happens with probability  $q_0^2 ( 1 - R )^2$.
  In this situation, function node $a$ is satisfiable by $j$ and $k$.
  
\item[(iii).]
  One of the two variable nodes, say $j$, is
  type-II unfrozen while the other, $k$, is type-I unfrozen.
  This happens with probability  $ 2 q_0^2  R ( 1 - R )$.
  In this situation, node $a$ is satisfiable by $j$.
  (It is also satisfiable by node $k$ through setting
  $\sigma_k = \sigma_k^*$, but this spin flipping will perturb
  the spin values of $O(N)$ other unfrozen variable nodes.)
  
\item[(iv).]
  Both node $j$ and node $k$ are type-I unfrozen, but they can not take
  the spin value $\sigma_j=\sigma_j^*$ and 
  $\sigma_k=\sigma_k^*$ simultaneously in state $\beta$.
  This happens with probability   $ q_0^2 R^2 / 2$.
  In this situation, node $a$ is definitely satisfied by one of
  nodes $j$ and $k$ (in some configurations of $\beta$, node $j$
  satisfies $a$ while $\sigma_k = - \sigma_k^*$, and in the
  remaining configurations node $k$ satisfies $a$ while
  $\sigma_j= - \sigma_j^*$).
  
\item[(v)]
  Both node $j$ and node $k$ are type-I unfrozen and they can take
  the spin value $\sigma_j=\sigma_j^*$ and $\sigma_k=\sigma_k^*$ 
  simultaneously.   This happens with probability   $ q_0^2 R^2  /2$.
  In this situation, node $a$ is also satisfiable by $j$ and $k$.
  However, to satisfy  $a$ by node $j$ or $k$ will cause an extensive
  perturbation to the spin configuration,
  whereby the spin values of $O(N)$ other unfrozen variable nodes
  will eventually be fixed.
  
\item[(vi).]
  One of the variable nodes, say $j$, is type-I unfrozen while the other is
  frozen to $\sigma_k=-\sigma_k^*$. This happens with probability
  $2 q_0 q_{-} R$.
  In this situation, node $a$ is 
  satisfiable by node $j$, with the same consequence as in situation (v).
  
\item[(vii).]
  One of the variable nodes, say $j$, is type-II unfrozen while the other
  is frozen to $\sigma_k=-\sigma_k^*$. This happens with probability
  $ 2 q_0 q_- ( 1 - R)$.
  In this situation, node $a$ is satisfiable  by node $j$.
  
\item[(viii).]
  Node $j$ is frozen to $\sigma_j=-\sigma_j^*$ and node $k$ is frozen to
  $\sigma_k=-\sigma_k^*$. This happens with probability $q_{-}^2$.
  In this situation, node $a$ is satisfiable only if
  the spin value of variable node $i$ is set to $\sigma_i= J_a^i$.
\end{description}

\begin{figure}[t]
  \psfig{file=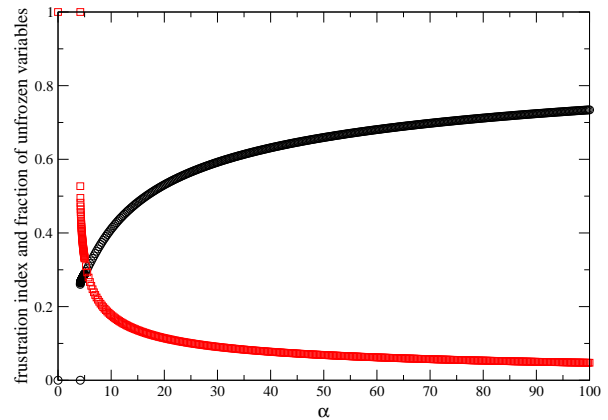,angle=270,width=10.0cm}
  \caption{ \label{fig:01}
    Long range frustration index $R$ (circles) and
    fraction of unfrozen variables $q_0$ (squares) as
    a  function of clause-to-variable ratio $\alpha$ for
    the random $3$-SAT problem.
  } 
\end{figure}

For the random $3$-SAT problem, the parameter $q_0$ is also determined
self-consistently by Eq.~(\ref{eq:06}), with the only difference
that $\lambda_1 = (3/2) \alpha [ q_0^2 R^2 /2 + 2 q_0 q_{-} R ]$ and
$\lambda_2 = (3/2) \alpha q_{-}^2$; and the parameter $R$ is
determined by Eq.~(\ref{eq:03}), with 
$\lambda_3 = 3 \alpha q_0 q_{-} (1 - R)$.
The lower and upper bound of the global minimum energy density
can be obtained through Eqs.~(\ref{eq:08}) and (\ref{eq:09}), with
$K=3$.
The values of the long range frustration order parameter $R$ and the
fraction of unfrozen variable nodes  $q_0$ are shown in 
Fig.~\ref{fig:01} as a function of $\alpha$,
and the energy density lower and upper bounds are
shown in Fig.~\ref{fig:03}.

When $\alpha < \alpha_R(3)= 4.189724$, Eqs.~(\ref{eq:06}) and
(\ref{eq:03}) permit only a trivial solution of $q_0 \equiv 1$ and
$R \equiv 0$, with zero energy density. When
$\alpha \geq \alpha_R(3)$, this non-frustrated replica symmetric
solution is still locally stable.
However, at this parameter range, another stable self-consistent
solution appears. This nontrivial solution is long--range
frustrated, with its order parameter $R$ jumps from zero to 
$R=0.2605$ at $\alpha= \alpha_R(3)$, accompanied by a drop of 
$q_0$ from $q_0 =1$ to $q_0=0.5270$.
Interestingly, although there are jumps for the order parameters
$R$ and $q_0$ at $\alpha_R(3)$, the energy density of this
solution increases only {\em gradually} from zero as a function
of $\alpha$ (see Fig.~\ref{fig:03}).
This is an improvement over a previous mean field
solution. If long range frustration was not taken into
account, Eq.~(\ref{eq:06}) predicts a
nontrivial solution when $\alpha \geq 4.667$, which,
however, has an unphysical negative energy density as long as
$\alpha < 5.18$ \cite{MonassonR1996,MonassonR1997}. 

The present mean field solution predicts that,
in the parameter range of $\alpha \geq \alpha_R(3)$
a random $3$-SAT formula will be long--range frustrated and
unsatisfiable.
This predicted SAT-UNSAT transition point is
located between the rigorously known lower and upper 
bounds \cite{AchlioptasD2001}.
However, 
through the zero temperature cavity method and population dynamics,
it is found in Ref.~\cite{MezardM2002b}
that, when $3.921 < \alpha < 4.267$
the random $3$-SAT problem has another non-frustrated
replica symmetry breaking solution with zero energy and
positive structural entropy. The structural entropy of
this non-frustrated solution becomes negative when
$\alpha \geq 4.267$, and this point is regarded as the
SAT-UNSAT transition point of the random $3$-SAT problem.
Therefore, there is a clear discrepancy between the present
work and the result of Ref.~\cite{MezardM2002b}. 
Two reasons are conceivable for this discrepancy:

First, the treatment in this paper is essentially replica symmetric, with
only two order parameters $q_0$ and $R$. In the random $3$-SAT
problem, it is clear that replica symmetry breaking occurs before
the onset of long range frustration. This is different from what
happens in the random $2$-SAT problem of Sec.~\ref{sec:2}
and the vertex cover problem of Ref.~\cite{ZhouH2004b}.
In those two systems, replica symmetry breaking occurs concomitantly
with the onset of long range frustration.  
When replica symmetry is broken and there are many (macroscopic) states
of global minimum energy, first, one needs more order parameters
to characterize the system in the non-frustrated phase; and second,
in the long--range frustrated phase,
the group of type-I unfrozen
variable nodes of one state as well as their
spin value correlations may be quite different from those of another
state. These two points make calculation quite difficult.
It is hoped that the present work will stimulate future efforts
on this line.

Second, the long--range frustrated mean field solution does
not evolve from the instability of the non-frustrated solution.
This is clear from the fact that, at $\alpha = \alpha_R(3)$,
the long range frustration order
parameter jumps to a positive value rather than gradually increases
from zero.
This is qualitatively different from what happens in the 
random $2$-SAT and the vertex cover problem, where long range
frustration is caused by a propagation and percolation of
frustration.  Maybe the present frustrated solution correspond to
a metastable macroscopic state
rather than a state of global minimum energy.
We believe that long range frustration will also exist
in the ground--states of the random $3$-SAT problem when
$\alpha$ is larger than some threshold value. If this onset of
long range frustration is caused be a percolation of frustration
as assumed in this paper, the order parameter $R$ should gradually
increases from zero. 

The author has also performed a stability
analysis of the replica symmetry breaking solution of 
Ref.~\cite{MezardM2002b}, to investigate whether or not
long range frustration will occur gradually when $\alpha \geq 4.267$.
His preliminary simulation results indicate that this solution is
also locally stable at $\alpha \sim 4.267$.

                                %
                                %

\section{
  Conclusion and discussion
  \label{sec:4}
}

In a random $K$-SAT problem, some of the variable nodes are unfrozen, 
their spin values change in different microscopic configurations
of global minimum energy. 
However, these fluctuations of spin values among
different unfrozen variable nodes may be highly correlated.
In this paper, we have investigated the possibility
of such kind of correlations based on the physical picture of percolation
transition.
For the random $2$-SAT problem, we found that
long range frustration among unfrozen variable nodes gradually
builds up  as the clauses-to-variables ratio $\alpha$ exceeds
$\alpha_R(2) = 4.4588$. This value is larger than the SAT-UNSAT transition 
point of $\alpha_c(2)=1$.
For the random $3$-SAT problem, we found a long--range
frustrated mean field solution when $\alpha > \alpha_R(3)=4.1897$.
The energy density of this solution increases gradually
from zero as a function of $\alpha$, while there is
a jump in the long range frustration order parameter $R$ 
at $\alpha_R(3)$. 
The SAT-UNSAT transition point of this nontrivial solution
is lower than the value of $\alpha_c(3)=4.267$ as predicted by
the work of Refs.~\cite{MezardM2002,MezardM2002b,BraunsteinA2002}.
Two possible reasons for this discrepancy were mentioned.
This discrepancy indicates that, when replica symmetry breaking
occurs before the onset of long range frustration, the present
mean field theory needs to be improved. We hope the present
work will stimulate further efforts to overcome this difficulty.

The phase diagram of the finite connectivity $\pm J$ 
Viana--Bray spin glass model
was also discussed. We suggested the possibility of a phase transition
from a non-frustrated spin glass phase to a long--range frustrated
spin glass phase as a finite temperature. This prediction might be
checked by numerical simulations.

The method used in the present work may also be applicable
to other finite connectivity spin glass models \cite{VianaL1985}. 
We hope that an appropriate combination
of long range frustration and the cavity method at the
first order replica symmetry breaking level \cite{MezardM2001,MezardM2003}
will help to advance our understanding of spin glass statistical
physics. 

There exists a rigorously polynomial algorithm to solve the $2$-SAT
decision problem; but the $3$-SAT decision problem is NP-complete.
The proliferation of macroscopic domains in the solution spaces of the
$3$-SAT problem and its absence in the $2$-SAT problem is unlikely to
be the reason for this qualitative difference in computational
complexity, since there exist 
class P (polynomial) combinatorial optimization problems
which also have exponentially many macroscopic states
in their solution spaces \cite{MezardM2003b,ZhouH2003d}.
In the author's opinion, an instance of a combinatorial
optimization problem may become intrinsically difficult if long range
frustration of the type discussed in this paper exists in
such a system.  Long range frustration causes strong correlations
among the fluctuations of the spin values of a finite fraction of 
the vertices of the system. This means that, if the spin 
on an unfrozen vertex is set to a given value, the spins on
other $O(N)$ unfrozen vertices should also be set to
the corresponding values. This spin value fixation process
may be extremely difficult to achieve even for a global
algorithm, since the algorithm needs to choose the
right vertices and the right spin values to fix. 

In the random $2$-SAT problem, there is no long range frustration
at both sides of the SAT-UNSAT transition point. In the 
vertex cover problem, there is an onset of long range
frustration in the computationally difficult case of
mean vertex degree $c \geq 2.7183$ \cite{ZhouH2004b}.
In the random $3$-SAT problem, we know there exists
a non-trivial solution with a jump of the long range
frustration order parameter $R$ at $\alpha_R(3)$. 
Although $\alpha_R(3)$ is unlikely to be true SAT-UNSAT
phase transition point for this problem, we expect to
observe that, in a correct replica symmetry breaking
mean  field solution long range frustration will
builds up when the clauses-to-variables ratio
$\alpha \geq \alpha_c(3)$. To construct such
a mean field solution is still an open problem.

                                %
                                %

\section*{
  Acknowledgement
}

The author is grateful to Reinhard Lipowsky and Lu Yu for support and
discussion. He also thanks Jing Han and Ming Li for their general
interest.

                                %
                                %


\end{document}